# Dual Synchronous Generator: Inertial Current Source based Grid-Forming Solution for VSC

Huanhai Xin, Kehao Zhuang, Pengfei Hu*, Yunjie Gu, Ping Ju

*Abstract-* In order to improve dynamic characteristics of the power system with high-proportion renewable energy sources (RESs), it is necessary for the voltage source converter (VSC), interfaces of RESs, to provide inertial and frequency regulation. In practical applications, VSCs are better to be controlled as a current source due to its weak overcurrent capacity. According to the characteristic, a dual synchronous theory is proposed to analyze the synchronization between current sources in this paper. Based on dual synchronous idea, a dual synchronous generator (DSG) control is applied in VSC to form inertial current source. In addition, a braking control is embedded in DSG control to improve the transient stability of VSC. Finally, experimental results verify the effectiveness of the theory and the control method.

## I. INTRODUCTION

Voltage source converters (VSCs), serving as the grid interfaces of renewable resources, are emerging as dominant elements in power systems. In line with this ongoing trend, VSCs are given increasing responsibilities of supporting grid frequency and voltage [1], which is characterized by the widely accepted terminology of grid-forming converters [1]. It is a common belief that grid-forming converters should somehow emulate the behavior of synchronous generators (SGs) in the sense of direct voltage control and rotating inertia [3]. However, VSCs are very sensitive to overloading due to the intrinsic limit of the fast-switchable power semiconductors, which implies that VSCs have to switch to current control mode during faults and therefore lose the expected grid-forming capacity. Such mode switches greatly compromised the benefits of grid forming converters and may even induce new transient stability problems [4], [5]. More importantly, the hard mode switching transients give rise to an ill-posed problem mathematically due to the restructuring of state variables and the network topology. As a result, most of the analytical techniques, including eigen analysis and Lyapunov method, become inapplicable. It is possible to use exhaustive numerical simulation to validate the transient stability with mode switching, but this is intractable for large-scale systems with volatile operation trajectories due to the curse of dimensionality.

For the reasons given above, it is highly desirable and necessary to develop grid-forming converters that have consistent behavior in normal operation and during faults with no hard mode switching. One possible way is to oversize the power semiconductors (and the associated heatsinks and passive components) to endow VSCs with very high short-circuit levels so that they behave as voltage sources during transients. This solution will significantly increase cost and ultimately set back the efforts of de-carbonization.

In this letter, we look for an alternative approach for consistent grid-forming control. It has been stated in Thevenin-Norton theorems that the independent sources in a circuit can be represented in the form of either voltage or current sources, and there are no apparent justifications that grid-forming capacities must be associated with voltage sources. From this observation, [6] and [7] establish a duality theory and point out that voltage and current sources share a unified mechanism of synchronization and power balancing. Based on the duality idea, this paper extends our previous work of virtual power angle control for grid-following converters [5] and proposes a current-based grid-forming control which is named Dual Synchronous Generator (DSG) to highlight its duality relationship to conventional SGs. The DSG is controlled as a rotating current source whose phase and frequency are governed by inertial-like dynamics taking feedbacks from the power at the terminal. The DSG can sustain an islanded grid by itself (and therefore is grid forming) and can synchronize to an infinite bus and other DSGs in a similar (or more precisely, dual) way to SGs. The DSG is a current source in both normal operation and faults, therefore it has consistent behavior in transients such that the hard mode switch between the current and voltage mode is avoided. It provides a new way for grid-forming control that is more suitable to the physical limits of VSCs and allows for most of the analytical methods used for whole-system analysis. It also points out a new direction to rethink the theoretical foundation of power system dynamics.

The rest of this letter is organized as follows. Section II introduces the principle of DSG. Characteristics of DSG are analyzed in Section III. Experimental results are presented in Section IV, which verify the feasibility of the proposed DSG. Section V concludes this letter.

## II. PRINCIPLE OF DUAL SYNCHRONOUS GENERATOR

### A. Duality theory of DSG and SG

The principle of DSG and its relationship with SG is illustrated in Fig. 1. Fig 1 (a) shows the duality representation of an infinite bus as either a voltage source or a current source according to Thevenin-Norton theorems. Thus are formed the corresponding single SG-DSG infinite bus system in Fig 1 (b). It has long been known that the dynamics of a SG system is governed by its power-angle relationship. We now show that the dynamics of a DSG system is governed by a similar principle.

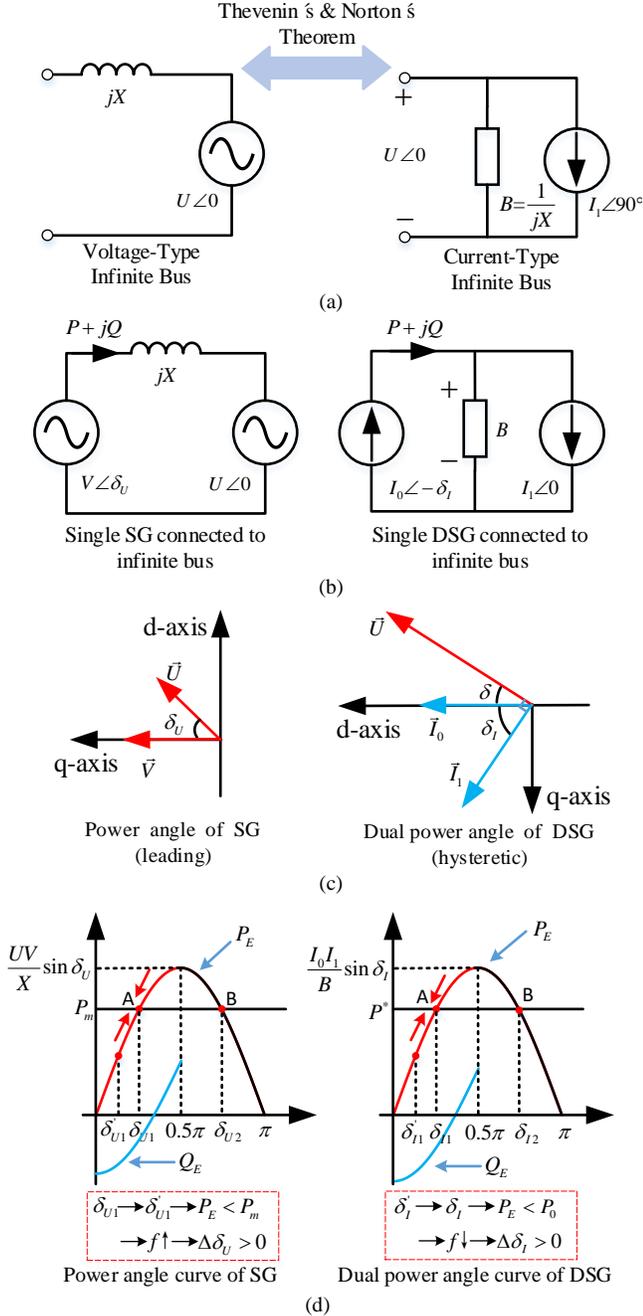

Fig.1 Duality of DSG and SG. (a) Representation of an infinite bus by equivalent voltage and current sources. (b) Equivalent circuit of single SG-DSG connected to an infinite bus. (c) Definition of power angle and dual power angle. (d) Power angle curves.

The transmitted power of the DSG system in Fig. 1 (b) can be calculated by

$$\begin{aligned} P_E + jQ_E &= \vec{V} \cdot (I_0 \angle \delta)^* \\ &= (U \angle 0 + jXI_0 \angle \delta) \cdot I_0 \angle -\delta \\ &= UI_0 \cos \delta_U + j(-UI_0 \sin \delta_U + I_0^2 X) \\ &= \frac{I_1 I_0}{B} \sin \delta_I + j\left(\frac{I_0^2}{B} - \frac{I_1 I_0}{B} \cos \delta_I\right) \end{aligned} \quad (1)$$

where $X=X_g+X_l$ denotes the impedance between the infinite bus and the LCL filter, $B$ equals $1/X$, $I_1$ denotes the grid current in Norton's equivalent circuit, with positive direction in Fig.1 (b), $I_0$ denotes the current source generated by VSC, $\delta_I$ named Dual Power Angle (DPA), denotes the angle between current source $I_0$ and $I_1$. Since $\mathbf{I_1} = -\mathbf{U}/(jX)$, the relationship between the conventional power angle $\delta_U$ and DPA $\delta_I$ satisfies $\delta_I = -\delta_U + (\pi/2)$.

The power-angle relationship in equation (1) is visualized as curves in Fig.1 (d), where two equilibrium points (A and B) exist. From (1), the DSG power equation is the same as that of synchronous generator. In addition, DPA is the angle of the hysteretic current phase of the VSC, which is dual to the leading power angle of SG. In order to ensure VSC have normal power characteristic which is same as synchronous generator, the P-f relationship is also dual to SG because of the duality of the power angle, the stable equilibrium is point A, as analyzed in Fig.1 (d). The maximum power $P_m=UI_{max}=(I_1 I_{max})/B$, where $I_{max}$ is decided by the current limitation of VSC. When the active power reference $P^*$ is larger than $P_E$, the frequency of VSC will decrease. When the active power reference $P^*$ is smaller than $P_E$, the frequency of VSC will increase. Therefore, normal operation range of DPA is $0 \sim \pi/2$.

According to the above analysis, the synchronization of DSG can be regarded as the synchronization between current sources. This dual form makes the synchronization analysis method of DSG exactly the same as that of synchronous generator and prevents power-angle curve switching of VSG under disturbance because of current limitation.

Table I summarizes the duality theory.

TABLE I
THE CHARACTERISTICS OF MASTER UNIT AND SLAVE UNIT

|  | DSG | SG |
|---|---|---|
| Synchronous | Synchronization between current sources | Synchronization between voltage sources |
| Power angle | Hysteresis angle | Leading angle |
| Active power expression | $P_E=(I_1 I_0 \sin\delta_I)/B$ | $P_E=(UV\sin\delta_U)/X$ |
| Reactive power expression | $Q_E=I_0^2/B-(I_1 I_0\cos\delta_I)/B$ | $Q_E=V^2/X-(UV\cos\delta_U)/X$ |
| Swing Characteristics | $f=G(s)(P_E-P^*)$ | $f=G(s)(P^*-P_E)$ |

B. The control strategy of DSG

Inspired by the idea in the dual synchronous stability theory, single-VSC-infinite-bus system shown in Fig.2 is used to analyze the DSG below. The proposed DSG control strategy block diagram is also illustrated in Fig.2 in detail. Considering the current limitation of VSC, it could be controlled as a current source, while the phase angle is oriented to the d-axis current. The synchronization loop is also illustrated in Fig.2, where a braking loop is introduced to a *P-θ* control where *J* and *D* emulates virtual inertia and damping ratio which is dual to the synchronous generator. As well, a proportional regulator ($K_P$) can be used in the synchronization loop, which acts as a SG without inertia. Meanwhile, the braking loop activated signal is generated by the reactive power, which will be analyzed comprehensively below. Another important control part is Q/V loop, which builds up relationships between the reactive power or magnitude of AC voltage and the d-axis current reference. This part can be either reactive power or AC voltage feedback control, while the transfer functions $G_V$ and $G_Q$ can be proportional regulator (*K*) or inertial link $1/(J_q s+D_q)$. Thus, this control structure could ensure the magnitudes of voltage and current fluctuate in a small range, while the q-axis current reference is set to zero, which could guarantee that the synchronous rotating frame is oriented to the current phase angle.

In order to analyze the proposed method easily, we first analyze the control strategy without the braking loop. The synchronous loop uses proportional control as an example, which can be expressed as

$$\omega - \omega_0 = K_P(P_E - P^*) \qquad (2)$$

where $\omega$ and $\omega_0$ are the angular frequencies of the converter and the grid respectively, $K_P$ is the P-f droop gain, $P_E$ and $P^*$ are the actual value and reference value of active power, respectively.

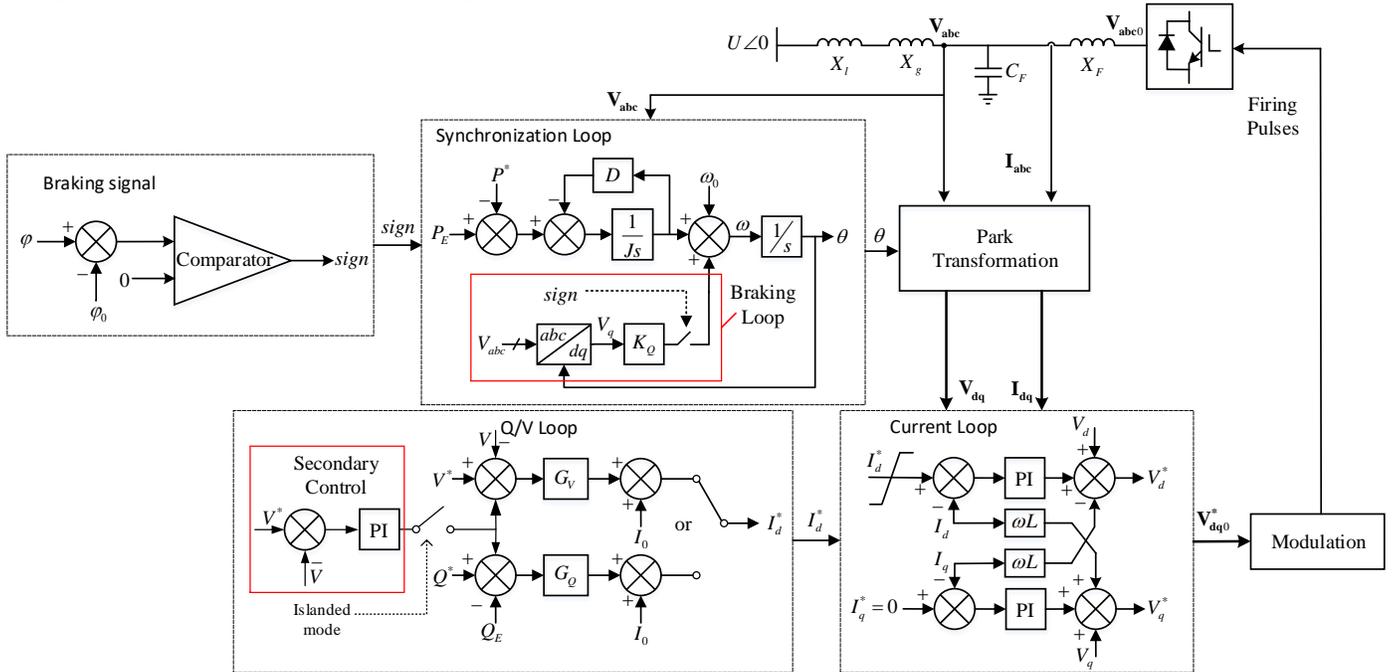

Fig.2 Dual Synchronous Generator control strategy for VSC

Taking V-I droop as an example, which can be expressed as

$$I_d^* - I_0 = K_V(V^* - V) \qquad (3)$$

where $I_0$ is the current RMS value, $I_d^*$ is the d-axis current reference, $K_V$ is the V-I droop gain, $V^*$ and $V$ are voltage reference value and actual value of RMS AC voltage.

V-I (or Q-I) control loop can ensure the constant output current of DSG, which is dual to the excitation control of SG. According to the Fig1 (c), the equivalent current of grid and DSG output current are constant, the active power is only related to DPA. The output power is changed by adjusting the DPA. Therefore, the adjustment of DSG power is also dual to SG.

When DSG operates in islanded mode, secondary control needs to be put into operation in order to keep the voltage in a reasonable range. However, This does not mean that DSG is a voltage source in islanded mode because secondary control is only a rough control.

### III. CHARACTERISTIC ANALYSIS OF TRANSIENT PROCESS

Based on the DPA curve in Fig.3 (a), there still exists transient stability issue under the condition of large voltage dip. There are two types of transient cases, i.e., the equilibrium point exists or equilibrium point not exists. As shown in Fig.3 (a), when the voltage dip is not serious, the equilibrium point D exists, which ensures the stability of the DSG method; when large voltage dip occurs (shown as the red line), transient instability will occur. Thus, a braking control loop is designed to avoid this unexpected Loss of Synchronization (LOS). From Fig.3 (a), it can be seen that the instability occurs when the operating point exceeds point F, which means that the maximum power point could not reach the power reference.

According to (1), the power factor of DSG is satisfies

$$\tan\varphi = \frac{Q_E}{P_E} = \frac{I_1 \cos\delta_I + I_0}{I_1 \sin\delta_I} \qquad (4)$$

where $\varphi$ denotes the power factor of DSG. $\varphi$ will increase with the increase of DPA. It can be seen from Fig.1 (d) and (4) that when $\delta_I=\pi$, $\varphi$ reaches $\pi/2$. Thus, according to (4), the following criteria is easily to be obtained: when $\varphi$ is smaller than limit angle $\varphi_0$, the braking loop is activated. The corresponding control strategy is shown in Fig.2, which can be expressed as follows

$$\omega - \omega_0 = K_P(P_E - P^*) + sign \cdot K_Q V_q \quad (5)$$

$$sign = \text{sgn}(\arctan\frac{Q_E}{P_E} - \varphi_0) \quad (6)$$

where $K_Q$ is the synchronization enhanced coefficient and $\vec{V_q}$ is the q-axis component of $\vec{V}$, function sgn(x) denotes 1 if x≥0, else 0. After activating sign(x) and substituting $\delta_I=\omega-\omega_0$ in to (5), it turns to

$$\dot{\delta}_I = K_P(P_E - P^*) + K_Q V_q = K_P(S - P^*) \quad (7)$$

$$S = P_E + \frac{K_Q}{K_P}V_q \quad (8)$$

where $S$ is the revised power. According to the circuit relationship, $V_q$ can be expressed as

$$V_q = U_q + XI_d \quad (9)$$

Meanwhile, $U_q$ can be expressed as

$$U_q = -U\cos\delta_I = I_1 X\cos\delta_I \quad (10)$$

Substituting (9), (10) into (8), yields

$$S = (P_m \sin\delta_I + \frac{K_Q I_1 X \cos\delta_I}{K_P}) + \frac{K_Q I_d X}{K_P}$$
$$= P_m' \sin(\delta_I + \alpha) + \frac{K_Q I_{max} X}{K_P} \quad (11)$$

where $P_m=I_0I_1/B$,

$$P_m' = \sqrt{P_m^2 + (K_Q I_1 X/K_P)^2}$$
$$\alpha = \arctan(K_Q I_1 X/K_P P_m) \quad (12)$$

$$K_{PL} = \frac{K_Q}{K_P}$$

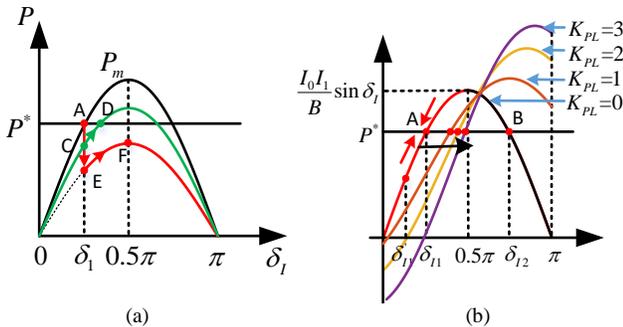

(a)      (b)
Fig.3 Operating trajectory of VSC under different disturbances (a) without braking loop (b) with braking loop

According to (11) and (12), the DPA curve is illustrated in Fig.3 (b), with $K_Q$ varying from 0 to $3K_P$. It can be seen from Fig.3 (b) that the braking loop could increase the maximum value of the DPA curve, improving the transient stability margin, since the instability issue occurs when the maximum value of DPA curve is smaller than $P^*$. Moreover, with the increasing of $K_Q$, the maximum value of DPA curve gets larger. Furthermore, according to (11), the maximum value of $S$ can be expressed as follow.

$$S_{max} = \sqrt{P_m^2 + (K_Q U/K_P)^2} + \frac{K_Q I_{max} X}{K_P} \quad (13)$$

If $S_{max}>P^*$ always holds, the equilibrium point will always exist. Thus, combining (13) and $S_{max}>P^*$ yields

$$\sqrt{P_m^2 + (K_Q U/K_P)^2} > P^* - \frac{K_Q I_{max} X}{K_P} \quad (14)$$

Thus, if $P^* - K_Q I_{max} X/K_P < 0$, i.e., $K_Q/K_P > P^*/(I_{max} X)$, holds, equation (14) will always hold, no matter the value of $I_1 X$. Hence, the criteria of the braking loop to maintain the transient stability is as follow.

$$K_Q/K_P > P^*/(I_{max} X) \quad (15)$$

IV. EXPERIMENTAL RESULTS

In order to verify the proposed DSG control method, hardware in the loop experiments of gird-connected are carried out. The experimental parameters are listed in Table II.

TABLE II
PARAMETERS OF VSC-GRID CONNECTED SYSTEM

| | |
|---|---|
| Base value of frequency for per-unit calculation $f_{base}$ | 50Hz |
| Base value of voltage for per-unit calculation $V_{base}$ | 380V |
| Base value of power for per-unit calculation $P_{base}$ | 10kVA |
| Grid-side inductance of line $X_{line}$ | 0.2p.u. |
| Grid-side inductance of LCL filter $X_g$ | 0.06p.u. |
| Converter-side inductance of LCL filter $X_F$ | 0.05p.u. |
| Capacitor of LCL filter $C_F$ | 0.05p.u. |
| P-f droop coefficient $K_p$ | 0.02 |
| Braking loop coefficient $K_Q$ | 0.06 |
| V-I droop coefficient $K_V$ | 0.01 |
| Current-loop proportional and integral coefficients $K_{ccP}$ $K_{ccI}$ | 0.4, 15 |

Fig.4 gives the time response of DSG at active power steps ($P^*$=0.3p.u.→0.5p.u.→0.7p.u.→1.0p.u.→1.05p.u.). When active power changes, the current of VSC is basically constant and the DSG voltage has slight fluctuation. With the increase of active power, reactive power increase gradually. It illustrates that rationality of duality theory and the great tracking performance of the control method.

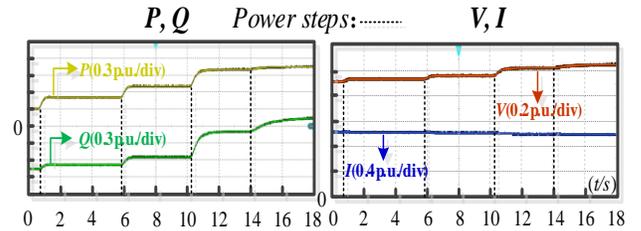

Fig.4 Response of DSG at active power step

Considering different voltage-sag conditions (U=1.0p.u.→ 0.8p.u. → 1.0p.u., U=1.0p.u. → 0.6p.u. → 1.0p.u.), the DSG transient performance with braking loop and without braking loop at $P^*$=0.8 p.u. is compared in Fig.5 and Fig.6. In Fig.5, when U drops to 0.8p.u., maximum active power $P_m$ is larger than active power reference $P^*$, both with braking loop and without braking loop are stable. In Fig.6, When $U$ drops to 0.6p.u., maximum active power $P_m$ is less than active power reference $P^*$. In Fig.6(a), the transient instability of the VSC without braking loop occurs under large disturbance. Fig.6(b) proves that the proposed method with braking loop can ensure the VSC ride-through transient occasions with no mode switching.

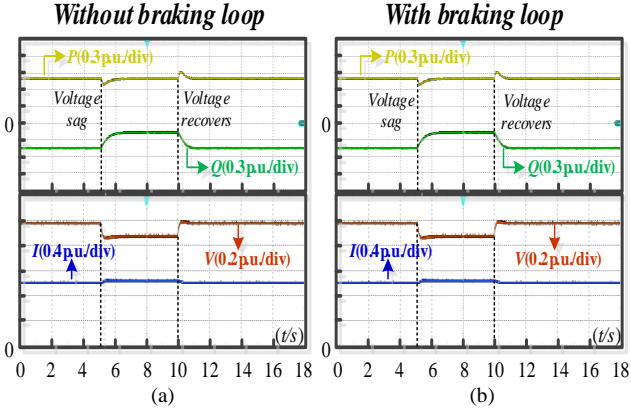

Fig.5 Response of DSG at U drops to 0.8p.u. (a) without braking loop (b) with braking loop

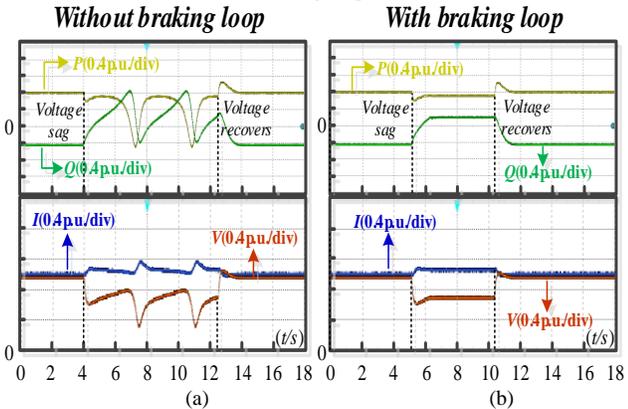

Fig.6 Response of DSG at U drops to 0.6p.u. (a) without braking loop (b) with braking loop

Fig.7 gives transient waveform of VSC with stability enhanced P-f control (SEPFC) [12] at $P^*$=0.7p.u.. From Fig.7, although VSC can maintain transient stability after mode switching (due to current saturation) at U drops to 0.6p.u., VSC cannot restore to normal mode (current unsaturation mode) when $U$ restores to 1p.u. It explains the disadvantage of VSC working as a voltage source.

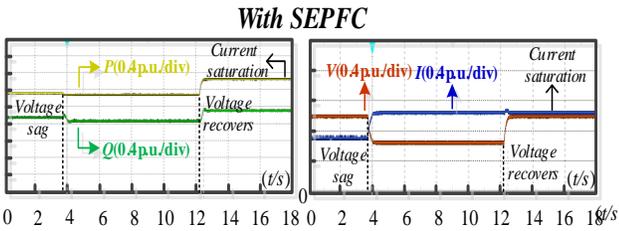

Fig.7 Response of SEPFC-inverter at U drops to 0.6p.u

In addition, in order to further explain the capability of grid-forming of DSG, Fig.8 gives the time response of a single DSG with load (1+j0.2 p.u.). The DSG maintains the current and voltage near 1 pu. The secondary control of DSG can ensure the voltage in an acceptable range. In island operation, the frequency of DSG can still maintain 50 Hz. Dual to SG, DSG can operate in island by forming the grid current.

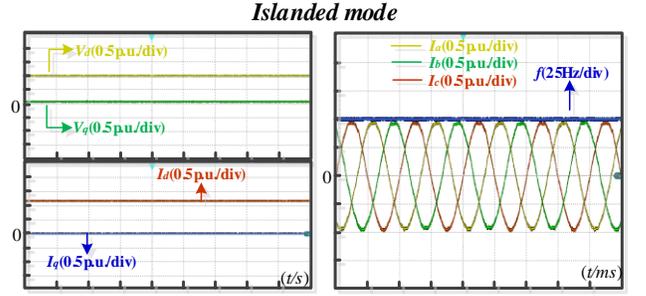

Fig.8 Response of DSG in islanded mode

## V. CONCLUSION

This letter proposes a dual synchronous generator based control method for grid-connected VSCs, which controls a VSC as a current source, unlike the existing grid-forming control methods controlling it as a voltage source. The control method forms the theory of dual synchronization, which reveals the essence of VSC synchronization. Based on the dual synchronization, a novel control strategy which could ensure both the transient stability and inertial provision capability of a grid-connected VSC is proposed, and the characteristic analysis of the control method is comprehensively presented as well. Finally, the experimental results validate the effectiveness of the proposed method, and comparisons with an existing method are conducted, proving the superiority of the proposed method.